\documentclass[12pt,preprint,amsmath,amssymb]{revtex4}

\usepackage{graphicx}%
\usepackage{dcolumn}%
\usepackage{bm}%

\begin{document}

\title{Quantum dynamics in macrosystems with several coupled electronic states: hierarchy of effective Hamiltonians}
\author{Etienne Gindensperger}
\author{Lorenz S. Cederbaum}
\affiliation{Theoretische Chemie, Universit{\"a}t Heidelberg, Im Neuenheimer Feld 229, D-69120 Heidelberg, Germany.}

\date{\today}

\begin{abstract}
We address the nonadiabatic quantum dynamics of macrosystems with several coupled electronic states, taking into account the possibility of multi-state conical intersections. The general situation of an arbitrary number of states and arbitrary number of nuclear degrees of freedom (modes) is considered. The macrosystem is decomposed into a system part carrying a few, strongly coupled modes, and an environment, comprising the vast number of remaining modes. By successively transforming the modes of the environment, a hierarchy of effective Hamiltonians for the environment is constructed. Each effective Hamiltonian depends on a reduced number of effective modes, which carry cumulative effects. By considering the system's Hamiltonian along with a few members of the hierarchy, it is shown mathematically by a moment analysis that the quantum dynamics of the entire macrosystem can be numerically exactly computed on a given time-scale. The time scale wanted defines the number of effective Hamiltonians to be included. The contribution of the environment to the quantum dynamics of the macrosystem translates into a sequential coupling of effective modes. The wavefunction of the macrosystem is known in the full space of modes, allowing for the evaluation of observables such as the time-dependent individual excitation along modes of interest, as well a spectra and electronic-population dynamics. 
\end{abstract}

\maketitle

\section{Introduction}

Within the last twenty years, so-called conical intersections have slowly emerged as a paradigm for ultrafast nonadiabatic processes occurring in polyatomic systems \cite{koppel:1984,michl:1990,yarkony:1996,domcke:1997,robb:2000,baer:2002,domcke:2004,worth:2004}. Conical intersections are widely spread, actually omnipresent, types of intersecting potential-energy surfaces, leading to a complete breakdown of the Born-Oppenheimer approximation. The electronic and nuclear motions are strongly coupled in the vicinity of the intersection, which thus provides a very efficient pathway for the ultrafast decay of the excited electronic state(s), typically on the femtosecond time scale.

Most of the studies reported to date exhibit intersections involving two electronic states. A notable exception is the case of triply-degenerate states due to symmetry requirement, as found, for instance, in the methyl cation \cite{1980.Katr.259}. More recent contributions also highlight accidental, {\em i.e.}, occurring at nonsymmetric geometries, simultaneous degeneracy of three electronic states. The latter has been found, for instance, in the ethyl \cite{2002.Mats.6907}, allyl \cite{2003.Mats.10672} and pyrazolyl \cite{2003.Mats.12428} radicals, in cytosine \cite{2004.Blan.10609} and in malonaldehyde \cite{2005.Coe.4560,2006.Coe.618}. Within a given electronic-state manifold, cases of intersections between different pairs of states have also been reported, for instance, in C$_2$H  \cite{2001.Mebe.3673} and in the benzene radical cation \cite{2006.Bald.064101}. Of course, two-state and three-state intersections can be present within the same manifold of electronic states, see, e.g. \cite{2005.Coe.4560}.

In this work, we aim at studying the multi-state non-adiabatic quantum dynamics in large polyatomic systems, {\em i.e.}, those involving a large number of nuclear degrees of freedom (modes). We refer to them as "macrosystems". These include, for instance, an impurity in a solid, a chromophore in a protein pocket, or a large isolated molecular system. The abundance of conical intersections grows with the dimensionality of the system and the density of electronic state in the energy domain considered. In large polyatomic systems, two-state intersections are known to be more common than avoided crossing situations \cite{truhlar:2003}. This argument, together with the fact that three-state intersections occur already in relatively small molecular species, let us suspect that multi-state intersections should be rather common and that several states are very likely to strongly interact nonadiabatically in macrosystems. We shall provide a theoretical framework to study the quantum dynamics in such situations. A special, but particularly important situation is given by the multi-mode Jahn-Teller effect \cite{englman,bersuker}.

Often, the dynamics of macrosystems can be viewed to be dominated by a "system" part comprising a few strongly coupled modes only. Then, the large number of remaining modes is seen as an "environment". This environment may play an important role by modifying the quantum dynamics provided by the system alone. Indeed, the environmental impact on the system's dynamics has been highlighted in numerous situations, and is particularly important in conical intersection situations due to the high sensitivity of the latter to even small perturbations \cite{1998.Wort.3518,1999.Raab.936,2002.Kuhl.263,burghardt:2004,prl:2005,paper2,molphys}. The impact of the environment should therefore be included in a realistic treatment of the macrosystem's dynamics. 

Nowadays, very powerfull methods exist to treat numerically exactly the quantum dynamics of molecular systems of moderate size. We think, in particular, of the {\em multiconfiguration time-dependent Hartree} (MCTDH) method \cite{meyer:1990,manthe:1992,beck:2000,mey03:251}, and to its multi-layer extension (ML-MCTDH) \cite{2003.Meye.251,2003.Wang.1289}. MCTDH is able to treat the quantum dynamics of systems involving 20-30 modes in two-state-intersection situations and of somewhat less modes if more states are involved. When even more modes are involved, as is typically the case in macrosystems, one natural approach is to use approximate quantum dynamical schemes which can account for conical intersection situations, see, for instance, Refs.~\cite{2005.Stoc.243,2006.Mart.119}. However, the full, {\em numerically exact}, quantum treatment of the dynamics is out of reach.

Recently, another strategy was proposed. The aim is to construct reduced models which can account for, at least, the dominant effects of the environment upon the system. The limited number of modes in such models allows for a numerically exact treatment of the quantum dynamics. Thus, quantum aspects such as interferences, geometric phase effects \cite{1992.Mead.51}, etc, are naturally taken into account. Such an approach has been proposed for two-state conical intersections in macrosystems \cite{prl:2005}. There, it has been shown that the use of three {\em effective} environmental modes only --together with the system's modes-- suffice to calculate accurately the band shape and short-time dynamics of the entire macrosystem. Detailled analysis of this effective-mode theory along with numerical applications can be found in Refs.~\cite{paper1,molphys,paper2,physica}. Precursors of this approach were derived more than twenty years ago for the Jahn-Teller effect \cite{obrien:1972,englman:1978,obrien:1980,fletcher:1980,cederbaum:1980,haller:1980}. This approach allows to split the environment into two parts: (i) a primary set of three effective modes which couples to the system's modes and carry the environmental effect on a short-time scale, and (ii) a "residual environment" which couples only to the effective modes and becomes important at later times. Importantly, the three effective modes are constructed from all the modes of the environment and carry the cumulative effects of the latter on the short-time scale. If time scales beyond short-times are under interest, the use of these three effective modes is, however, not sufficient, and one has to take into account the residual environment. This can be done by constructing additional sets of effective modes, as recently proposed in Ref.~\cite{triplet}. Based on extensive numerical examples, it has been highlighted that the systematic use of additional effective modes allows to calculate accurately the quantum dynamics for longer and longer times. In this vein, a related extension of the effective-mode theory has been used to analyze exciton dissociation in semiconducting polymers \cite{irene,hep}.

In this theoretical paper, we extend these recent findings to the general case of multi-state intersections in macrosystems, and analyze the dynamical properties of the proposed approach.
We shall provide a detailled scheme for the construction of successive sets of effective modes in the situation where any number of electronic states are coupled. More specifically, we consider the general case of an arbitrary number of coupled electronic states and an arbitrary number of environmental modes. It will be shown mathematically that the use of the system's modes augmented by a limited number of sets of effective modes suffice to calculate, numerically exactly, the quantum dynamics of the entire macrosystem on a given time-scale.

The paper is constructed as follows. In section II, we discuss the Hamiltonian used to describe the macrosystem with several coupled electronic states. In section III, we detail the construction of the sets of effective modes, which allows us to decompose the Hamiltonian of the environment into a hierarchy of effective Hamiltonians. In section IV, the dynamical properties of this hierarchy are analyzed and discussed. Section V concludes.

\section{The Hamiltonian of the macrosystem}

We start with the Hamiltonian $\hat{H}$ of the macrosystem. $\hat{H}$ describes $N$ coupled electronic states in a diabatic representation. From now on, the "hat" symbol refers to $N \times N$ matrices in the electronic space. We define $\hat{H}$ as follows:
\begin{equation}
\label{eq:H}\hat{H}=\hat{H}_S+\hat{H}_B,
\end{equation}
with a "system" Hamiltonian given by
\begin{equation}
\label{eq:HS}\hat{H}_S=\{h^S_{i,j}\},\quad i,j =1,\ldots,N,
\end{equation} where the elements $h^S_{i,j}$ are Hamiltonians depending on the $N_S$ nuclear degrees of freedom (modes) of the system. These Hamiltonians $h^S_{i,j}$ are not further specified here, since all the developments which will follow do not depend on their particular form. Apart form containing the kinetic energy operators for the system's modes, they can, for instance, include the full (diabatic) potential energy surfaces obtained from ab-initio data for these modes. In general, $\hat{H}_S$ can couple all the $N$ electronic states. In this work, the rest of the Hamiltonian, $\hat{H}_B$, which we shall call "Hamiltonian of the environment", is described by the linear vibronic coupling (LVC) model \cite{koppel:1984}. This model corresponds to the first term in a Taylor expansion of the actual potential energy surfaces at a given point in configuration space (reference geometry), in terms of so-called normal modes. This well established model has been successfully used to treat the quantum dynamics of molecular systems involving several coupled electronic states, see, for instance, Refs~\cite{koppel:1984,domcke:2004,worth:2004}. The Hamiltonian $\hat{H}_B$ is thus given by:
\begin{equation}
\label{eq:HB}\hat{H}_B=h_0\hat{1}+ \{h^B_{i,j}\},
\end{equation} with
\begin{eqnarray}
\label{eq:H0} h_0 &=& \sum_{k=1}^{N_B} \frac{\omega_k}{2} (p_k^2+x_k^2),\\
\label{eq:HBij}h^B_{i,j} &=& \sum_{k=1}^{N_B} \kappa_k^{(i,j)} x_k,\quad \mathrm{for}\quad i,j=1,\ldots,N
\end{eqnarray} where $N_B$ is the number of modes of the environment and $\hat{1}$ the unit matrix in the electronic space. The $x_k$ are position operators and the $p_k$ the corresponding momentum operators expressed using mass- and frequency-weighted quantities. The diagonal elements of $\hat{H}_B$ consist of harmonic oscillators, $h_0$, augmented by state- and mode-dependent shifts given by $h^B_{i,i}$. The off-diagonal element $h^B_{i,j}$ couples the electronic states $i$ and $j$. All the $N_B$ modes of the environment can, in principle, couple all the electronic states.

Equations (\ref{eq:H}) to (\ref{eq:HBij}) determine the Hamiltonian of our macrosystem which contains $N_S+N_B$ modes and involves $N$ coupled electronic states. The above definition of the Hamiltonian of the macrosystem constitutes a natural extension of the 2-state Hamiltonian used as starting point in closely related former works on conical intersections in macrosystems \cite{prl:2005,paper1,paper2,molphys}. 

Let us discuss the Hamiltonian introduced above for $N$ coupled electronic states. In Eq.~(\ref{eq:H}) we have isolated a "system" part from the rest ("environment") of the Hamiltonian. This separation is not always needed, but is usefull at least in two important situations. (i) When we aim at studying the impact of an environment on a "system", it is obviously convenient to use the proposed form of the total Hamiltonian. (ii) When some modes of the macrosystem (presumably a few, see below) cannot be reasonably treated with the LVC model, they can be treated as being part of a "system", which, in our approach, is {\em not} restricted to any kind of model. Of course, if all the modes of the macrosystem can be satisfactorily described by the LVC model, we can include all of them in $\hat{H}_B$. In this case, $\hat{H}_S$ does not describe any dynamics and reduces to a diagonal matrix with the energy of the electronic states at the chosen reference geometry as elements. Note that the modes entering $H_S$ and $H_B$ do not couple directly, but do couple indirectly via the electronic subsystem. This is easily recognized by noticing that $\{h^B_{i,j}\}$ in Eq.~(\ref{eq:HB}) provides an explicit system-environment interaction.  

We make some further remarks on the LVC model. First, this model is valid for symmetry enforced as well as accidental intersections of potential energy surfaces. In this work, we even allow for the possibility that each environmental mode may contribute simultaneously to the diagonal as well as off-diagonal parts of the Hamiltonian. This allows one to account for cases where the environment lacks any symmetry. Of course, the LVC model can also describe all standard special situations, among which is the Jahn-Teller effect \cite{englman,bersuker}. The LVC model can be thought of as the minimal general model for the dynamics of intersecting potential energy surfaces.

We suppose throughout this paper that the number of modes of the system is not too large, so that one can compute the quantum dynamics provided by $H_S$. It is the potentially very large number of environmental modes which renders the computation of the quantum dynamics of the entire macrosystem impossible. In the following, we shall introduce systematic transformations of the $N_B$ modes of the environment which will allow us to exactly decompose $\hat{H}_B$ into a hierarchy of effective Hamiltonians. These effective Hamiltonians, which will depend on a reduced number of effective modes, will be shown in Sec IV to translate into a sequential description of the dynamics of the entire macrosystem. Thus, depending on the time-scale under interest, only the first members of the hierarchy of effective Hamiltonians will be needed to compute the dynamics of the entire macrosystem. Since each effective Hamiltonian will depend only on a reduced number of effective modes, the dynamics can be computed numerically (if, of course, the number of effective Hamiltonians taken into account is not too large.) This approach has been recently derived for the 2-state case and shown by a numerical example to produce very accurate results for the quantum dynamics, see Ref.~\cite{triplet} (see also a related perspective in Refs.~\cite{irene,hep}). In the following sections we construct the hierarchy of effective Hamiltonians for $N$ coupled electronic states and study its dynamical properties.

\section{Construction of the hierarchy of effective Hamiltonians}

We shall introduce successive orthonormal transformations of the modes of the environment which will allow us to build the hierarchy of effective Hamiltonians. The system part of the total Hamiltonian will not be affected by these transformations of the environmental modes. The first step in the construction of the hierarchy will allow us to split $\hat{H}_B$ into two parts:
\begin{equation}
\label{eq:H1Hr1} \hat{H}_B = \hat{H}_1 + \hat{H}_{r1}.
\end{equation}
The first part, $\hat{H}_1$, which constitutes the first member of the hierarchy of effective Hamiltonians, will be shown to take account of {\em all} the couplings between the electronic states due to the environment and will thus play a key role. Importantly, $\hat{H}_1$ will depend only on a limited set of effective modes, constructed from all the modes of the environment. The second or remaining part of $\hat{H}_B$, called $\hat{H}_{r1}$, will be {\em diagonal} in the electronic space. 

In a second step, we will iteratively construct additional sets of effective modes out of the remaining modes of $\hat{H}_{r1}$ and obtain a hierarchy of effective Hamiltonians $\hat{H}_{m}$:
\begin{eqnarray}
\nonumber \hat{H}_{r1} &=& \hat{H}_{2} + \hat{H}_{r2}\\ 
\nonumber \hat{H}_{r2} &=& \hat{H}_{3} + \hat{H}_{r3}\\
\nonumber &\vdots&\\
\hat{H}_{rm} &=& \hat{H}_{m+1} + \hat{H}_{rm+1},
\end{eqnarray} where, in each step, the next member of the hierarchy, $\hat{H}_{m+1}$, is constructed from the former remaining part $\hat{H}_{rm}$.

\subsection{Construction of the first effective Hamiltonian}

We construct in the following the first member of the hierarchy of effective Hamiltonians, $\hat{H}_1$. We shall give a detailed scheme for this construction; a similar scheme will be used in the subsequent construction of the rest of the hierarchy. The content of this subsection can be viewed as an extention of the results for the 2-state case \cite{prl:2005,paper1} to the general $N$-state case of interest in this paper. To construct the first member of the hierarchy, we shall identify effective modes in $\hat{H}_B$, and introduce accordingly an appropriate orthonormal transformation of all the $N_B$ modes of the environment. 

\subsubsection{Introduction of the effective modes}

Inspecting $\hat{H}_B$, we see that we can define {\em effective} modes. Since $\hat{H}_B$ is hermitian, there exists a maximal number of $\mathcal{N}=N(N+1)/2$ independent elements $h^B_{i,j}$ for $i,j=1,\ldots,N$, see Eqs.~(\ref{eq:HB}) and (\ref{eq:HBij}). We write them in terms of $\mathcal{N}$ effective modes $\tilde{X}_l$, $l=1,\ldots,\mathcal{N}$:
\begin{equation}
\label{eq:tildeX} h^B_{i,j}=\bar{\kappa}^{(i,j)}\tilde{X}_{l},\quad l=(i-1)N+j-i+1,
\end{equation}
with {\em effective coupling constants} $\bar{\kappa}^{(i,j)}$ defined by
\begin{equation}
\label{eq:bark}\bar{\kappa}^{(i,j)} = \left( \sum_{k=1}^{N_B} (\kappa^{(i,j)}_k)^2 \right)^{1/2}.
\end{equation}
We assume that the number of environmental modes, $N_B$, is larger than $\mathcal{N}$; if it is not the case our approach is not usefull as such. 

It is convenient to use a matrix- and vector-notation. We introduce the column vector of the $N_B$ initial modes $\bm{x}=(x_1,\ldots,x_{N_B})^T$, the vector of the $\mathcal{N}$ new modes, $\tilde{\bm{X}}_{\mathcal{N}}=(\tilde{X}_1,\ldots,\tilde{X}_{\mathcal{N}})^T$, and a matrix $\bm{V}_{\mathcal{N} \times N_B}$. The bold-faced quantities indicate vectors and matrices in the configuration space, and the subscripts indicate the size of the matrix or vector (no subscript means the full space, $N_B$). We define $\bm{V}_{\mathcal{N} \times N_B}$ as: 
\begin{equation}
\label{eq:V}\bm{V}_{\mathcal{N} \times N_B} = \left( \begin{array}{ccc}
\frac{\kappa_1^{(1,1)}}{\bar{\kappa}^{(1,1)}}   & \cdots & \frac{\kappa_{N_B}^{(1,1)}}{\bar{\kappa}^{(1,1)}} \\
\frac{\kappa_1^{(1,2)}}{\bar{\kappa}^{(1,2)}}   & \cdots &  \frac{\kappa_{N_B}^{(1,2)}}{\bar{\kappa}^{(1,2)}}\\
\vdots & & \vdots\\
\frac{\kappa_1^{(N-1,N)}}{\bar{\kappa}^{(N-1,N)}} & \cdots & \frac{\kappa_{N_B}^{(N-1,N)}}{\bar{\kappa}^{(N-1,N)}} \\
\frac{\kappa_1^{(N,N)}}{\bar{\kappa}^{(N,N)}}   & \cdots & \frac{\kappa_{N_B}^{(N,N)}}{\bar{\kappa}^{(N,N)}} \end{array} \right).
\end{equation}
We straightforwardly have
\begin{equation}
\label{eq:x1}\tilde{\bm{X}}_{\mathcal{N}}=\bm{V}_{\mathcal{N} \times N_B}\bm{x}.
\end{equation}

The $\tilde{X}_l$, $l=1,\ldots,\mathcal{N}$, are already normalized, but are {\em not} orthogonal to each other. We orthogonalize them by using a standard method, e.g., the Gram-Schmidt orthogonalization procedure, and call the corresponding orthogonal modes ${X}_l$, $l=1,\ldots,\mathcal{N}$ and the $\mathcal{N}$-dimensional vector they form $\bm{X}_{\mathcal{N}}=(X_1,\ldots,X_{\mathcal{N}})^T$. We thus have
\begin{equation}
\label{eq:X} \bm{X}_{\mathcal{N}} = \bm{U}_{\mathcal{N} \times \mathcal{N}} \tilde{\bm{X}}_{\mathcal{N}},
\end{equation} with $\bm{U}_{\mathcal{N} \times \mathcal{N}}$ being a $\mathcal{N} \times \mathcal{N}$ matrix which orthogonalizes the modes $\tilde{X}_l$, $l=1,\ldots,\mathcal{N}$. Combining Eqs.~(\ref{eq:x1}) and (\ref{eq:X}) we arrive at
\begin{equation}
\label{eq:x2}\bm{X}_{\mathcal{N}} = \bm{T}_{\mathcal{N} \times N_B} \bm{x}
\end{equation}
where
\begin{equation}
\bm{T}_{\mathcal{N} \times N_B} = \bm{U}_{\mathcal{N} \times \mathcal{N}} \bm{V}_ {\mathcal{N} \times N_B}.
\end{equation}
This defines a {\em set of $\mathcal{N}$ orthonormal effective modes}, ${X}_l$, $l=1,\ldots,\mathcal{N}$, constructed from all the original modes $x_k$, $k=1,\ldots,N_B$.

By applying the transformation $\bm{T}_{\mathcal{N} \times N_B}$ to the original modes of the environment, the terms $h^B_{i,j}$ of $\hat{H}_B$ can be expressed as 
\begin{equation}
\label{eq:hbX}h^B_{i,j}=\bar{\kappa}^{(i,j)}\sum_{l=1}^{\mathcal{N}} K_l^{(i,j)} {X}_{l},
\end{equation}
with the coefficients $K_l^{(i,j)}$ given by
\begin{equation}
\label{eq:K}K_l^{(i,j)} = \sum_{k=1}^{N_B} \frac{\kappa_k^{(i,j)}}{\bar{\kappa}^{(i,j)}} t_{lk}, 
\end{equation} 
where the $t_{lk}$ are the elements of $\bm{T}_{\mathcal{N} \times N_B}$. Note that the $K_l^{(i,j)}$ satisfy the normalization conditions $\sum_{l=1}^{\mathcal{N}} (K_l^{(i,j)})^2 = 1\quad \forall i,j=1,\ldots,N$. The coefficients $K_l^{(i,j)}$ represent how the contributions to the effective coupling constant $\bar{\kappa}^{(i,j)}$ are distributed among the $\mathcal{N}$ effective modes.

In Eq.~(\ref{eq:tildeX}), each $h^B_{i,j}$ depends on a single (non-orthogonal) effective mode. After the orthonormalization of these modes, each element $h^B_{i,j}$ can depend on {\em all} the $\mathcal{N}$ orthonormal effective modes defined in Eq.~(\ref{eq:x2}). Importantly, by applying the transformation $\bm{T}_{\mathcal{N} \times N_B}$, we have transformed Eq.~(\ref{eq:HBij}) into Eq.~(\ref{eq:hbX}): the $h^B_{i,j}$ depend now only on the $\mathcal{N}$ effective modes, instead of the $N_B$ original modes. It is already evident now that these effective modes will play a crucial role in the dynamics of the macrosystem, since they suffice to represent all the coupling terms $h^B_{i,j}$ present in the Hamiltonian of the environment.

We have constructed, so far, $\mathcal{N}$ effective modes which we call the primary effective modes. The full Hamiltonian of the environment contains $N_B$ modes and, therefore, in order to fully describe $H_B$ we need to identify $N_B-\mathcal{N}$ additional modes. This amounts to construct a full, $N_B$-dimensional orthogonal transformation matrix. We call this matrix $\bm{T}^{(1)}$. The superscript $(1)$ means that this is the first transformation related to the construction of $\hat{H}_{1}+\hat{H}_{r1}$, see Eq.~(\ref{eq:H1Hr1}), and that subsequent transformations will be used later when constructing the other members of the hierarchy. Defining the complete set of new modes as $\bm{X}=(X_1,\ldots,X_{N_B})^T$, the matrix $\bm{T}^{(1)}$ is obviously such that   
\begin{equation}
\label{eq:x}\bm{X} = \bm{T}^{(1)} \bm{x}.
\end{equation}
In fact, it is easily recognized that the matrix $\bm{T}_{\mathcal{N} \times N_B}$ contains the $\mathcal{N}$ first rows of the complete orthonormal transformation matrix $\bm{T}^{(1)}$. These rows determine the primary effective modes. The remaining $N_B-\mathcal{N}$ rows of $\bm{T}^{(1)}$ can be chosen in many ways as long as $\bm{T}^{(1)}$ is orthonormal (see Sec. III.A.2 for a particularly appealing choice). Obviously, the transformation in Eq.(\ref{eq:x}) leads to the elements $h^B_{i,j}$ given by Eq.~(\ref{eq:hbX}). To fully transform the total Hamiltonian of the environment, we have to transform the diagonal part $h_0\hat{1}$ as well, see Eq.~(\ref{eq:HB}). This gives
\begin{equation}
\label{eq:H0X}h_0 = \sum_{l=1}^{N_B} \frac{\Omega_l}{2} (P_l^2+X_l^2) + \sum_{l<l'=1}^{N_B} d_{ll'} (P_l P_{l'} + X_l X_{l'}),
\end{equation} with $P_l$ the momentum associated with $X_l$ and
\begin{equation}
\label{eq:Omega}\Omega_l = \sum_{k=1}^{N_B} \omega_k t^{2}_{lk},\quad\quad d_{ll'}=\sum_{k=1}^{N_B} \omega_k t_{lk}t_{l'k},
\end{equation}  where the $t_{lk}$ are the elements of full transformation matrix $\bm{T}^{(1)}$. Eqs.~(\ref{eq:hbX}) and (\ref{eq:H0X}) give $\hat{H}_B$ in the new, complete set of modes.

We are now in the position to isolate the part of the transformed $\hat{H}_B$ which contains only the $\mathcal{N}$ primary effective modes and to obtain the desired form of $\hat{H}_B$
\begin{equation}
\label{eq:HB1r1}\hat{H}_B = \hat{H}_1+\hat{H}_{r1},
\end{equation}
where the Hamiltonian $\hat{H}_1$ containing only the $\mathcal{N}$ effective modes is the first member of the hierarchy and reads
\begin{eqnarray} 
\nonumber\label{eq:H1}\hat{H}_1 &=& \sum_{l=1}^{\mathcal{N}} \frac{\Omega_l}{2} (P_l^2+X_l^2)\hat{1} + \{\bar{\kappa}^{(i,j)}\sum_{l=1}^{\mathcal{N}} K_l^{(i,j)} {X}_{l}\}\\
&&+ \sum_{l<l'=1}^{\mathcal{N}} d_{ll'} (P_l P_{l'} + X_l X_{l'})\hat{1}.
\end{eqnarray}
The residual part $\hat{H}_{r1}$ takes on the diagonal form in the electronic space
\begin{eqnarray}
\nonumber \label{eq:Hr1}\hat{H}_{r1} &=&  \sum_{l=\mathcal{N}+1}^{N_B} \frac{\Omega_l}{2} (P_l^2+X_l^2)\hat{1}\\
\nonumber &&+ \sum_{l<l'=\mathcal{N}+1}^{N_B} d_{ll'} (P_l P_{l'} + X_l X_{l'})\hat{1}\\
&&+ \sum_{l=1}^{\mathcal{N}}\sum_{l'=\mathcal{N}+1}^{N_B} d_{ll'} (P_l P_{l'} + X_l X_{l'})\hat{1}.
\end{eqnarray}
It is worth noting that since the transformation is complete and orthonormal, the new form of $\hat{H}_B$ in Eq.~(\ref{eq:HB1r1}) is completely equivalent to the original one given by Eq.~(\ref{eq:HB}): these two versions of the Hamiltonian describe obviously exactly the same physical problem. In the new version of the Hamiltonian, $\hat{H}_1$ depends on three terms. The first one consists of harmonic oscillators, the second one couples the electronic states, and the third one contains additional bilinear kinetic and potential terms which couple the $\mathcal{N}$ primary effective modes among themselves. The remaining part, $\hat{H}_{r1}$, also contains three terms. The first (harmonic oscillators) and second (bilinear kinetic and potential coupling terms) ones depend only on the $N_B-\mathcal{N}$ remaining modes. Importantly, none of these modes participates in the coupling between the electronic states! The third term is made of bilinear kinetic and potential couplings between the $\mathcal{N}$ primary effective modes and the remaining modes. All terms of $\hat{H}_{r1}$ are diagonal in the electronic space. Thus, the transformation $\bm{T}^{(1)}$ decomposes $H_B$ into two parts, one, $\hat{H}_1$, which does couple the electronic states and one, $\hat{H}_{r1}$, which does not. $\hat{H}_{1}$ depends on $\mathcal{N}$ effective modes only. For the 2-state case, we recover that we need three effective modes \cite{prl:2005,paper1}. For the 3-state case, 6 effective modes and for the 4-state problem, 10 effective modes are needed, and so on. We remind that due to specific properties of some macrosystems, {\em e.g.}, those of high symmetry, one may need less than $\mathcal{N}$ primary effective modes to construct $\hat{H}_{1}$, but never more. The number of effective modes is indeed equal to the number of linearly independent elements $h_{i,j}^B$ in the original Hamiltonian of the environment $\hat{H}_B$, see Eq.~(\ref{eq:HBij}). Various special cases of the 2-state case have been discussed in Ref.~\cite{paper1}, and can now be easily extended to the present context. Among others, a particularly important case is the multi-state multi-mode Jahn-Teller problem. 

\subsubsection{A unique choice for the orientation of the effective modes}

To derive $\hat{H}_{B}=\hat{H}_{1}+\hat{H}_{r1}$ it was unnecessary to explicitly specify two "quantities": (i) the particular choice of the orthogonalization matrix $\bm{U}_{\mathcal{N} \times \mathcal{N}}$, see Eq.~(\ref{eq:X}), and (ii) the choice of the $N_B-\mathcal{N}$ remaining rows of the full orthonormal transformation matrix $\bm{T}^{(1)}$ in Eq.~(\ref{eq:x2}). The final results for $\hat{H}_{1}$ and $\hat{H}_{r1}$ in Eqs.~(\ref{eq:H1})-(\ref{eq:Hr1}) are valid whatever these choices for these quantities are. The freedom of choice is related to (i) the orientation of the $\mathcal{N}$ primary effective modes which span $\hat{H}_1$ on one hand, and (ii) the orientation of the $N_B-\mathcal{N}$ remaining modes on the other hand.

All choices of orientation {\em within} the two subspaces of modes leads to mathematically equivalent results. However, some particular choices can lead to usefull simplifications of $\hat{H}_1$ and $\hat{H}_{r1}$. These simplifications concern the bilinear kinetic and potential coupling terms among the effective modes within $\hat{H}_1$ on the one hand, and among all the modes within $\hat{H}_{r1}$ on the other hand. We want to stress that these simplifications are not needed to construct the remaining members of the hierarchy (see below), but give rise to working equations in a more closed form. Moreover, this form is particularly amenable to interpretation.

We now introduce a particular orientation of the $\mathcal{N}$ modes entering $\hat{H}_{1}$. Indeed, it is possible to exactly remove the third term of Eq.~(\ref{eq:H1}), {\em i.e.}, to remove {\em all} the bilinear coupling terms {\em among} the $\mathcal{N}$ primary effective modes. To proceed, we first notice that once the $\mathcal{N}$ modes have been constructed as detailled above, one can rotate them among themselves without changing the space they span, and thus without altering the impact of $\hat{H}_{1}$. The particular orientation of interest is the one which diagonalizes the $\mathcal{N} \times \mathcal{N}$ matrix with elements $\sum_{k=1}^{N_B} \omega_k t_{lk} t_{l'k}$, $l,l'=1,\ldots,\mathcal{N}$, see Eq.~(\ref{eq:Omega}). The eigenvalues of this matrix are now the frequencies of the rotated modes, and the set of eigenvectors provides the $\mathcal{N} \times \mathcal{N}$ rotation matrix from the initial set of effective modes to the new, rotated one. The procedure is fully described in Appendix E of Ref.~\cite{paper1} for the 2-state case, and is easily extended to the $N$-state situation discussed in this paper. This procedure corresponds to a {\em unique} choice of the orientation of the $\mathcal{N}$ primary modes, or, equivalently, to a unique choice of the orthogonalization matrix $\bm{U}_{\mathcal{N} \times \mathcal{N}}$, see Eq.~(\ref{eq:X}).

The same procedure, but now for the remaining $N_B-\mathcal{N}$ modes, exactly removes the second term of $\hat{H}_{r1}$ in Eq.~(\ref{eq:Hr1}) by appropriately choosing the orientation of these modes. This procedure fixes the remaining $N_B-\mathcal{N}$ rows of the transformation $\bm{T}^{(1)}$ in Eq.~(\ref{eq:x}). As a result of this procedure the Hamiltonian takes on the simplified form 
\begin{eqnarray}
\label{eq:HB1r1v2}\hat{H}_B &=& \hat{H}_1+\hat{H}_{r1},\\
\label{eq:H1v2}\hat{H}_1 &=& \sum_{l=1}^{\mathcal{N}} \frac{\Omega_l}{2} (P_l^2+X_l^2)\hat{1} + \{\bar{\kappa}^{(i,j)}\sum_{l=1}^{\mathcal{N}} K_l^{(i,j)} {X}_{l}\},\\
\nonumber \label{eq:Hr1v2}\hat{H}_{r1} &=&  \sum_{l=\mathcal{N}+1}^{N_B} \frac{\Omega_l}{2} (P_l^2+X_l^2)\hat{1}\\
&&+ \sum_{l=1}^{\mathcal{N}}\sum_{l'=\mathcal{N}+1}^{N_B} d_{ll'} (P_l P_{l'} + X_l X_{l'})\hat{1},
\end{eqnarray} where we have kept the same notation as in Eqs.~(\ref{eq:HB1r1})-(\ref{eq:Hr1}), although all quantities refer to the modes rotated as described above. Note that the $\mathcal{N}$ effective modes entering $\hat{H}_{1}$ are now only coupled through the electronic subsystem, and that in $\hat{H}_{r1}$ the remaining $N_B-\mathcal{N}$ modes are decoupled from each other. The equations (\ref{eq:HB1r1v2})-(\ref{eq:Hr1v2}) constitute our working equations.

Apart form the fact that the mathematical form of Eqs.~(\ref{eq:HB1r1v2})-(\ref{eq:Hr1v2}) is simpler than the one of Eqs.~(\ref{eq:HB1r1})-(\ref{eq:Hr1}), which is of interest by itself and has numerical advantages, the use of the above unique choice of orientation of the $\mathcal{N}$ effective modes and of the residual modes is motivated by the following physical arguments. The first argument concerns $\hat{H}_{1}$ which is the only part of $\hat{H}_{B}$ which couples the electronic states directly. In Eq.~(\ref{eq:H1}), the primary effective modes are coupled among themselves in two ways, indirectly {\em via} the electronic subspace, and directly by the bilinear kinetic and potential coupling terms. 
By using the proposed orientation, the direct couplings among the primary effective modes are eliminated. As a particular consequence, the kinetic energy operator of the modes entering $\hat{H}_{1}$ takes on the usual form, and all the "unpleasant" momentum-space couplings disappear. This has an important implication. Now, if the off-diagonal elements of the matrix $\hat{H}_{1}$ go to zero, {\em i.e.}, if there is no electronic coupling, the primary effective modes are no more coupled to each other as is the case in the usual picture of the LVC model. If, on the other hand, other choices of orientation are used, see also the related discussion for the 2-state situation in Refs.~\cite{paper1,molphys}, the modes remain coupled to each other.

The second argument concerns the residual part $\hat{H}_{r1}$, and is related to the distributions of the couplings between the primary effective modes and the residual modes as a function of the frequencies of the latter. Here, again, the unique orientation of the $N_B-\mathcal{N}$ residual modes chosen suppresses the bilinear couplings among them and leads to the usual form of the kinetic energy within the subspace of residual modes. This choice also leads to a unique set of $\mathcal{N}$ distributions of the bilinear couplings $d_{ll'}$ between the primary and residual effective modes (one distribution for each primary mode $l$, see Eq.~(\ref{eq:Hr1v2})). These distributions explicitly reflect {\em all} the coupling between the primary and residual effective modes as the residual modes are not coupled among themselves. This provides a particular physical meaning to the above distributions, in addition to the simpler mathematical form of the Hamiltonian obtained by the unique choice of modes. 

\subsection{Construction of higher members of the hierarchy}

In the preceding section we provided the first member of the hierarchy of effective Hamiltonians. We shall now pursue the construction of the hierarchy. This is done by iteratively transforming the $N_B-\mathcal{N}$ modes of $\hat{H}_{r1}$ which is the residual part of $\hat{H}_B$, given by Eq.~(\ref{eq:Hr1v2}). These successive transformations will allow us to decompose $\hat{H}_{r1}$ into a hierarchy of effective Hamiltonians $\hat{H}_{2}+\hat{H}_{3}+\ldots$ as follows:
\begin{equation}
\hat{H}_{rm} = \hat{H}_{m} + \hat{H}_{rm+1}.
\end{equation} The underlying idea is to construct additional sets of effective modes which successively carry cumulative effects of the respective residual part of the environment. The residual part $\hat{H}_{rm}$ of the environment shrinks as the hierarchy proceeds. This idea has been recently used in the construction of a hierarchy of effective Hamiltonians for the 2-state case in Ref.~\cite{triplet}. We also refer to the recent work of Tamura {\em et al.}~\cite{irene,hep} for a closely related construction of a hierarchy of 2-state effective Hamiltonians in the context of exciton dissociation in semiconducting polymers. The hierarchy of effective modes derived previously for the 2-state case is extended here to the $N$-state case.

\subsubsection{The second member of the hierarchy}

To construct $\hat{H}_{2}$ and $\hat{H}_{r2}$ out of $\hat{H}_{r1}$, we employ a very similar approach to the one used above to define $\hat{H}_{1}$  and $\hat{H}_{r1}$ out of $\hat{H}_{B}$. The approach consists of three steps:\\ 
(i) identify additional collective (effective) modes in $\hat{H}_{r1}$,\\
(ii) orthonormalize these effective modes,\\
(iii) rotate these orthonormalized modes to obtain a more convenient mathematical and physical form of the equations.\\
\noindent The first step is achieved by appropriately rewriting the term which couples $\hat{H}_{r1}$ to $\hat{H}_{1}$, {\em i.e.}, the last term of Eq.~(\ref{eq:Hr1v2}). We keep for the moment only the position-dependent part and write
\begin{eqnarray}
\label{eq:addmode}\sum_{l=1}^{\mathcal{N}}\sum_{l'=\mathcal{N}+1}^{N_B} d_{ll'} X_l X_{l'}=\sum_{l=1}^{\mathcal{N}} \bar{d}_l X_{l} \sum_{l'=\mathcal{N}+1}^{N_B} \frac{d_{ll'}}{\bar{d}_l} X_{l'},
\end{eqnarray} where we define the effective coupling constants $\bar{d}_l$ by
\begin{equation}
\label{eq:bard}\bar{d}_l^2 = \sum_{l'=\mathcal{N}+1}^{N_B} d_{ll'}^2,\quad \mathrm{for}\quad l=1,\ldots,\mathcal{N}.
\end{equation} 
In Eq.~(\ref{eq:addmode}) we readily recognize $\mathcal{N}$ additional, normalized, effective modes defined by $\tilde{X}_{\mathcal{N}+l} = \sum_{l'=\mathcal{N}+1}^{N_B} (d_{ll'}/\bar{d}_l) X_{l'}$, $l=1,\ldots,\mathcal{N}$. Equivalently, in a matrix notation:
\begin{equation}\label{eq:x2tilde}
(\tilde{X}_{\mathcal{N}+1},\ldots,\tilde{X}_{2\mathcal{N}})^T = \bm{V}^{(2)}_{\mathcal{N}\times(N_B-\mathcal{N})} (X_{\mathcal{N}+1},\ldots,X_{N_B-\mathcal{N}})^T
\end{equation}
with
\begin{equation}
\bm{V}^{(2)}_{\mathcal{N}\times(N_B-\mathcal{N})} = \left( \begin{array}{ccc}
d_{1,\mathcal{N}+1}/\bar{d}_1 & \cdots & d_{1,2\mathcal{N}}/\bar{d}_1 \\
\vdots & & \vdots \\
d_{\mathcal{N},\mathcal{N}+1}/\bar{d}_{\mathcal{N}} & \cdots & d_{\mathcal{N},2\mathcal{N}}/\bar{d}_{\mathcal{N}} \end{array} \right).
\end{equation}
The number of these additional modes is of course equal to the number of primary effective modes. As seen in Eq.~(\ref{eq:addmode}), these modes are coupled to the primary modes. They are obviously orthonormal to the $\mathcal{N}$ primary effective modes, but are not orthogonal among themselves. We orthogonalize them using an identical procedure as the one given in section III.A.1, and call these $\mathcal{N}$ orthonormalized modes $\tilde{\tilde{X}}_{k}$, $k=\mathcal{N}+1,\ldots,2\mathcal{N}$: 
\begin{equation} \label{eq:x3tilde}
(\tilde{\tilde{X}}_{\mathcal{N}+1},\ldots,\tilde{\tilde{X}}_{2\mathcal{N}})^T = \bm{U}^{(2)}_{\mathcal{N}\times\mathcal{N}} (\tilde{X}_{\mathcal{N}+1},\ldots,\tilde{X}_{N_B-\mathcal{N}})^T
\end{equation} with $\bm{U}^{(2)}_{\mathcal{N}\times\mathcal{N}}$ being the orthogonalization matrix. Using Eqs.~(\ref{eq:x2tilde}) and (\ref{eq:x3tilde}) we obtain
\begin{equation}
(\tilde{\tilde{X}}_{\mathcal{N}+1},\ldots,\tilde{\tilde{X}}_{2\mathcal{N}})^T =  
\bm{T}^{(2)}_{\mathcal{N}\times(N_B-\mathcal{N})}
(X_{\mathcal{N}+1},\ldots,X_{N_B-\mathcal{N}})^T
\end{equation} where 
\begin{equation}
\bm{T}^{(2)}_{\mathcal{N}\times(N_B-\mathcal{N})} = \bm{U}^{(2)}_{\mathcal{N}\times\mathcal{N}}
\bm{V}^{(2)}_{\mathcal{N}\times(N_B-\mathcal{N})}.
\end{equation}

The matrix $\bm{T}^{(2)}_{\mathcal{N}\times(N_B-\mathcal{N})}$ corresponds to the first $\mathcal{N}$ rows of a full, $(N_B-\mathcal{N}) \times (N_B-\mathcal{N})$, orthonormal transformation matrix $\bm{T}^{(2)}$. As it was the case for $\bm{T}^{(1)}$ in the construction of the first effective Hamiltonian, the remaining rows of $\bm{T}^{(2)}$ are chosen such as $\bm{T}^{(2)}$ is orthonormal. We finally obtain our vector of new orthonormalized effectives modes
\begin{equation}
(\tilde{\tilde{X}}_{\mathcal{N}+1},\ldots,\tilde{\tilde{X}}_{N_B})^T =  
\bm{T}^{(2)}
(X_{\mathcal{N}+1},\ldots,X_{N_B})^T.
\end{equation}
Transforming accordingly $\hat{H}_{r1}$, we obtain
\begin{equation}
\label{eq:Hr1v3}\hat{H}_{r1} = \hat{H}_{2} + \hat{H}_{r2}
\end{equation}
where $\hat{H}_{2}$ is the second member of the hierarchy of effective Hamiltonians and reads (we drop the double tilde on the modes for simplicity)
\begin{eqnarray}
\nonumber\label{eq:H2}\hat{H}_{2} &=& \sum_{k=\mathcal{N}+1}^{2\mathcal{N}} \frac{\Omega_{k}}{2} (P_{k}^2+X_{k}^2)\hat{1}\\
 && + \sum_{l=1}^{\mathcal{N}} \bar{d}_l \sum_{k=\mathcal{N}+1}^{2\mathcal{N}} A_{lk} (P_l P_{k} + X_l X_{k})\hat{1},
\end{eqnarray}
and the residual part now takes on the appearance
\begin{eqnarray}
\nonumber \label{eq:Hr2}\hat{H}_{r2} &=&  \sum_{k'=2\mathcal{N}+1}^{N_B} \frac{\Omega_{k'}}{2} (P_{k'}^2+X_{k'}^2)\hat{1}\\
 && + \sum_{k=\mathcal{N}+1}^{2\mathcal{N}} \sum_{k'=2\mathcal{N}+1}^{N_B} d_{kk'} (P_k P_{k'} + X_k X_{k'})\hat{1}.
\end{eqnarray}
Here, we have further used the unique choices of orientation of the $\mathcal{N}$ modes of $\hat{H}_{2}$, and of the corresponding residual $N_B-2\mathcal{N}$ modes of $\hat{H}_{r2}$ which simplify the equations as done in section III.A.2 for $\hat{H}_{1}$ and $\hat{H}_{r1}$. The procedure to determine these orientations is exactly the same. The quantities $A_{lk}$, $\Omega_{k}$ and $d_{kk'}$ are easily determined by the analogous equations to those for the quantities for $\hat{H}_{1}$, {\em i.e.}, Eq.~(\ref{eq:K}) and (\ref{eq:Omega}), respectively. 

We see from Eqs.~(\ref{eq:H2})-(\ref{eq:Hr2}) that only the $\mathcal{N}$ secondary effective modes entering $\hat{H}_{2}$ couple now to the $\mathcal{N}$ primary effective modes of $\hat{H}_{1}$, whereas all the $N_B - \mathcal{N}$ modes of $\hat{H}_{r1}$ do so before the transformation. The $N_B - 2\mathcal{N}$ remaining modes of $\hat{H}_{r2}$ couple to the secondary effective modes, and do not couple directly to the primary effective modes. Of course, since $\hat{H}_{r1}$ is diagonal in the electronic space, $\hat{H}_{2}$ and $\hat{H}_{r2}$ are also diagonal.

\subsubsection{The complete hierarchy}

Inspecting the new residual part $\hat{H}_{r2}$ in Eq.~(\ref{eq:Hr2}), we immediately see that it is exactly of the same mathematical form as $\hat{H}_{r1}$ in Eq.~(\ref{eq:Hr1v2}), except that $\hat{H}_{r2}$ depends only on the last $N_B - 2\mathcal{N}$ modes which are the residual modes of $\hat{H}_{1}+\hat{H}_{2}$. Consequently, we can use again our procedure described above to determine the third and higher members of the hierarchy of effective Hamiltonians. Eventually, this amounts to iteratively transforming the full Hamiltonian $\hat{H}_B$ of the environment to the sum
\begin{equation}
\label{eq:HBfinal}\hat{H}_{B}=\hat{H}_{1} + \sum_{m=2}^{N_{\mathcal{N}}} \hat{H}_{m},
\end{equation} where the primary effective Hamiltonian $\hat{H}_1$ is given by Eq.~(\ref{eq:H1v2}), and the higher members of the hierarchy $\hat{H}_{m}$ for $m=2,\ldots,N_{\mathcal{N}}$ all have the same formal structure and read
\begin{eqnarray}
\label{eq:Hm}\hat{H}_{m} &=&  \sum_{l'=(m-1)\mathcal{N}+1}^{m\mathcal{N}} \frac{\Omega_{l'}}{2} (P_{l'}^2+X_{l'}^2)\hat{1}\\
\nonumber &+& \sum_{l=(m-2)\mathcal{N}+1}^{(m-1)\mathcal{N}} \bar{d}_l \sum_{l'=(m-1)\mathcal{N}+1}^{m\mathcal{N}} A_{ll'} (P_l P_{l'} + X_l X_{l'})\hat{1}.
\end{eqnarray}
Except of $\hat{H}_1$, all the $\hat{H}_{m}$ are diagonal in the electronic space. The number of sets of $\mathcal{N}$ effective modes is given by $N_{\mathcal{N}}=$ {\em IntegerPart} $(N_B/\mathcal{N})$, with the integer part being here the smallest integer greater or equal to $N_B/\mathcal{N}$. To simplify the following discussion, we shall call each set of $\mathcal{N}$ effective modes a {\em multiplet of modes}. Note that the last multiplet may include less modes than $\mathcal{N}$. 

Thus, $\hat{H}_{B}$ contains a total of ${N_{\mathcal{N}}}$ multiplets of modes. $\hat{H}_1$, which couples the electronic states directly, depends on the first multiplet only. Each higher member of the hierarchy $\hat{H}_{m}$ couples to $\hat{H}_{m-1}$ and is diagonal in the electronic space. Consequently, while in the original $\hat{H}_{B}$ all the modes played formally the same role, after the successive transformations we get a {\em sequential coupling of the multiplets of modes}. 

If one includes all the members of the hierarchy in a calculation on the macrosystem, of course, the exact results are recovered since all modes of the full Hamiltonian are taken into account. Generally, this is not possible in practice for a macrosystem because the number of environmental modes, $N_B$, or, equivalently, the number of multiplets $N_{\mathcal{N}}$, can be very large. In any practical calculation, we will {\em truncate} the Hamiltonian of the macrosystem. In this respect, the new form of the Hamiltonian $\hat{H}_{B}$ of the environment given by Eq.~(\ref{eq:HBfinal}), allows us to truncate the environment in a very systematic manner by resorting to the highest number of effective Hamiltonians one can or wants to afford for the problem at hand. It is indeed evident that truncating the hierarchy in this way is on a completely different level of quality than simply neglecting some of the environmental modes in the initial $H_B$. Each member of the hierarchy of effective Hamiltonians is constructed from all the modes of the residual environment, and carries cumulative effects represented by the effective coupling constants $\bar{\kappa}$ and $\bar{d}$. Furthermore, only the first multiplet of modes couples directly the electronic states, while in the original $\hat{H}_{B}$ {\em all} the modes of the environment couple these states.

We shall reveal in the following the intimate relationship between the hierarchy of effective Hamiltonians and the dynamical properties of the full macrosystem. The accuracy achieved when using a truncated hierarchy for computing the quantum dynamics of the entire macrosystem comprised of an arbitrary number of coupled electronic states and of an arbitrary number of modes is studied in the next section.

\section{Properties of the hierarchy and moments analysis}

We shall show in the following that a truncated hierarchy of effective Hamiltonians allows us to compute {\em numerically exactly} the quantum dynamics of the {\em entire macrosystem} on a given time-scale. The number of effective Hamiltonians included in the calculation will be shown to provide the corresponding time-scale. By including more and more effective Hamiltonians, the dynamics can be numerically exactly computed on a longer and longer time-scale. This central result is proven here by using an analysis of the moments of the full macrosystem.

\subsection{Moments analysis}

We consider the autocorrelation function $P(t)$ of the full macrosystem
\begin{equation}
P(t)=\langle \underline{0} | e^{-i\hat{H}t} | \underline{0} \rangle,
\end{equation}
where $| \underline{0} \rangle$ is a column vector in the electronic space with components $\tau_i |0\rangle$, $i=1,...,N$. $|0\rangle$ is the initial nuclear wavefunction in the initial -- usually ground -- electronic state and the $\tau_i$ are the transition matrix elements between this initial state and the manifold of $N$ coupled electronic states. Depending on the experiment performed, $|\tau_i|^2$ is the oscillator strength or ionization cross section for the state $i$. We suppose troughout this paper that the $\tau_i$ are (complex-valued) constants, {\em i.e.}, are not position-dependent, and assume $\sum_{i=1}^{N} |\tau_i|^2 = 1$. We suppose also that the initial nuclear wavefunction $|0\rangle$ can be factorize as $|0_S \rangle |0_B \rangle$, where $|0_S \rangle$ is the initial wavefunction of the system and $|0_B \rangle$ the noninteracting ground-state wavefunction of the environment. Thus, $|0_B \rangle$ is separable with respect to the original environmental modes, and remains separable after our series of orthonormal transformations which lead to the final set of effective modes for the fully transformed $\hat{H}_B$. Note that this is true because we use mass- and frequency-weighted coordinates and momenta. For convenience, we write $|0_B \rangle$ as a direct product of nuclear wavefunctions for each multiplet of effective modes, rather than for the individual effective modes. Importantly, the initial wavefunction of the system, $|0_S \rangle$, is {\em not} restricted to any particular form, as is the case for the system's Hamiltonian $\hat{H}_S$. In particular, $|0_S \rangle$ can include nonseparable contributions with respect to the system's modes. Our initial nuclear wavefunction is thus given by
\begin{equation}
\label{eq:0}|0\rangle=|0_S \rangle|0_1\rangle|0_{2}\rangle\cdots|0_{N_\mathcal{N}}\rangle,
\end{equation} where the subscripts label the multiplets of environmental modes (and not the modes themselves), with
\begin{equation}
|0_m\rangle \propto exp \left(-\sum_{l=(m-1)\mathcal{N}+1}^{m\mathcal{N}} X_l^2\right)
\end{equation} for all $m=1,\ldots,\mathcal{N}$. We assume $|0\rangle$ to be normalized.

The autocorrelation function $P(t)$ measures the overlap between the initial wavefunction and that at later times $t$ \cite{1992.Mant.9062,1992.Enge.76}. The moments of the Hamiltonian are obtained by a Taylor expansion in time of the autocorrelation function \cite{kampen:1992}
\begin{equation}
P(t)=\sum_{n=0}^{\infty} \frac{(-it)^n}{n!} M_n,
\end{equation} where $M_n$ is the $n$-th order moment of the full macrosystem and is given by:
\begin{equation}
\label{eq:MM}M_n = \langle \underline{0} | \hat{H}^n | \underline{0} \rangle.
\end{equation}

As the time $t$ increases, more and more moments will contribute to the dynamical evolution of the macrosystem. The spectrum of the macrosystem corresponds to the Fourier transform of the autocorrelation function. Thus, the moments $M_n$ are also connected to the spectral properties of the macrosystem. $M_0$ is the total intensity, $M_1$ gives the energy location, $M_2$ is related to the width and $M_3$ to the main asymmetry of the spectrum, etc. It has been shown for the 2-state case that $\hat{H}_S+\hat{H_1}$ reproduces {\em exactly} the moments of the macrosystem up to and including $M_3$ \cite{prl:2005,paper1} and this has opened the field of effective modes for multi-mode conical intersections. Recently, the hierarchy of effective Hamiltonians has been introduced for the 2-state case \cite{triplet,irene}. We extend this result to the full hierarchy with $N$ electronic states and formulate a {\em theorem}:

{\em Consider a macrosystem with $N$ coupled electronic states described by the Hamiltonian $\hat{H}=\hat{H}_S+\hat{H}_B$ of Eq.~(\ref{eq:H}). The system's Hamiltonian augmented by the $n$ first members of the hierarchy of effective Hamiltonians, $\hat{H}_S + \sum_{m=1}^{n} \hat{H}_m$, suffices to reproduce exactly all the moments $M_k$ with $k \leq 2n+1$ of the entire macrosystem.}

Below we present a rigorous proof of this theorem. Before doing so, we mention that while writing this paper we learned that Tamura {\em et al.} \cite{hep} discussed the moments in the 2-state situation without considering a system part $\hat{H}_S$ and came to the same conclusion for this special case.

We start the proof with the fully transformed Hamiltonian, Eq.~(\ref{eq:HBfinal}), which contains $N_\mathcal{N}$ multiplets of effective modes build from all the $N_B$ modes of the environment, plus the $N_S$ modes of the system
\begin{equation}
\label{eq:H3}\hat{H} = \hat{H}_S + \sum_{m=1}^{N_\mathcal{N}} \hat{H}_m
\end{equation}
and rearrange as
\begin{equation}
\label{eq:mathH}\hat{H} = \sum_{m=1}^{N_\mathcal{N}} (\hat{\mathcal{H}}_m + \mathcal{E}_m \hat{1}),
\end{equation} where
\begin{equation} \label{eq:mathH1}\hat{\mathcal{H}}_1=\hat{H}_S+\hat{H}_{1}-\mathcal{E}_1\hat{1},\quad \hat{\mathcal{H}}_m = \hat{H}_m - \mathcal{E}_m\hat{1},\; \forall m\geq 2.\end{equation} 
$\mathcal{E}_m$ is the zero-point energy of the $m$-th multiplet of modes and we also introduce $\mathcal{E}=\sum_{m=1}^{N_\mathcal{N}} \mathcal{E}_m$, which corresponds to the zero-point energy of the full Hamiltonian of the environment. Note that, for convenience, we include the system part in $\hat{\mathcal{H}}_{1}$ in Eq.~(\ref{eq:mathH1}), {\em i.e.}, only $\hat{\mathcal{H}}_{1}$ couples the electronic states. We stress that the Hamiltonian $\hat{H}$ in Eq.~(\ref{eq:mathH}) is completely equivalent to the original Hamiltonian of the entire macrosystem given by Eq.~(\ref{eq:H}), since the entire hierarchy as well as the system are considered. The moments $M_n$ of the Hamiltonian $\hat{H}$ are thus exactly the moments of the entire macrosystem. From Eqs.~(\ref{eq:MM})-(\ref{eq:mathH}) we obtain
\begin{eqnarray}
M_n &=& \langle \underline{0} | \hat{H}^n | \underline{0} \rangle = \langle \underline{0} | (\sum_{m=1}^{N_\mathcal{N}} \hat{\mathcal{H}}_m + \mathcal{E}\hat{1})^n | \underline{0} \rangle\\
&=& \sum_{k=0}^{n} \binom{n}{k} \mathcal{E}^{n-k} \langle \underline{0} | (\sum_{m=1}^{N_\mathcal{N}} \hat{\mathcal{H}}_m)^k | \underline{0} \rangle,\\
\label{eq:Mn}&=&  \sum_{k=0}^{n} \binom{n}{k} \mathcal{E}^{n-k} \mathcal{M}_k,
\end{eqnarray}
where the coefficients $\binom{n}{k}$ are the usual binomial coefficients, and the partial moments $\mathcal{M}_k$ are given by
\begin{equation} 
\mathcal{M}_k = \langle \underline{0} | (\sum_{m=1}^{N_\mathcal{N}} \hat{\mathcal{H}}_m)^k | \underline{0} \rangle. 
\end{equation} 
To evaluate the moment $M_n$ we have to evaluate the partial moments $\mathcal{M}_k$ for all $k\leq n$. 

Before proceeding with the calculation of the moments, let us give two usefull properties of the $\hat{\mathcal{H}}_m$:
\begin{eqnarray}
&(P1)&\quad \hat{\mathcal{H}}_m |\underline{0} \rangle = 0\quad \forall m \geq 2\\
&(P2)&\quad \left[ \hat{\mathcal{H}}_k, \hat{\mathcal{H}}_l \right]  = 0\quad \mathrm{if}\quad l \neq k\pm1
\end{eqnarray}
The property (P1) is easily proven using Eqs.~(\ref{eq:mathH1}), (\ref{eq:Hm}) and (\ref{eq:0}). It is also straightforward to prove (P2): $\hat{\mathcal{H}}_k$ depends on the multiplets $k$ and $k-1$, and $\hat{\mathcal{H}}_l$ on the multiplets $l$ and $l-1$, see Eq.~(\ref{eq:Hm}). We readily see that, if $l\geq k+2$ or if $l\leq k-2$ then $\hat{\mathcal{H}}_k$ and $\hat{\mathcal{H}}_l$ do not depend on the same multiplets and thus commute. For $k=l$ it is trivial that they also commute. Recall that the $\hat{\mathcal{H}}_k$ are diagonal in the electronic space for all $k>1$. The two properties (P1) and (P2) follow directly from the particular form of the initial nuclear wavefunction and of the Hamiltonians which compose the hierarchy.

We return to the evaluation of the moment $M_n$ and evaluate the term $\mathcal{M}_k$ of Eq.~(\ref{eq:Mn}) separately for odd and even powers of $k$ which we denote for simplicity $2l$ and $2l+1$, respectively. We write
\begin{eqnarray}
\nonumber \label{eq:nn} \mathcal{M}_{2l} &=& \langle \underline{0} | (\sum_{m=1}^{N_\mathcal{N}} \hat{\mathcal{H}}_m)^l (\sum_{m=1}^{N_\mathcal{N}} \hat{\mathcal{H}}_m)^l | \underline{0} \rangle\\
\label{eq:M2n} &=& \langle \underline{l} |  \underline{l} \rangle\\
\nonumber \label{eq:nn+1} \mathcal{M}_{2l+1} &=& \langle \underline{0} | (\sum_{m=1}^{N_\mathcal{N}} \hat{\mathcal{H}}_m)^l (\sum_{m=1}^{N_\mathcal{N}} \hat{\mathcal{H}}_m)^{l+1} | \underline{0} \rangle\\
\label{eq:M2n+1}&=& \langle \underline{l} |  \underline{l+1} \rangle
\end{eqnarray}
where we define
\begin{equation}
\label{eq:n}|\underline{l} \rangle \equiv (\sum_{m=1}^{N_\mathcal{N}} \hat{\mathcal{H}}_m)^l | \underline{0} \rangle = \sum_{m=1}^{N_\mathcal{N}} \hat{\mathcal{H}}_m |  \underline{l-1} \rangle.
\end{equation}
Let us evaluate $|\underline{1} \rangle$. Making use of the property (P1), one immediately finds 
\begin{equation} \label{eq:1}|\underline{1} \rangle = \sum_{m=1}^{N_\mathcal{N}} \hat{\mathcal{H}}_m |\underline{0} \rangle = \hat{\mathcal{H}}_1 |\underline{0} \rangle.
\end{equation} 
The state $|\underline{2} \rangle$ reads
\begin{eqnarray}
|\underline{2} \rangle &=& \sum_{m=1}^{N_\mathcal{N}} \hat{\mathcal{H}}_m |\underline{1} \rangle = \sum_{m=1}^{N_\mathcal{N}} \hat{\mathcal{H}}_m \hat{\mathcal{H}}_1 |\underline{0} \rangle\\
\label{eq:2} &=& (\hat{\mathcal{H}}_1 + \hat{\mathcal{H}}_2) \hat{\mathcal{H}}_1 |\underline{0} \rangle, \end{eqnarray} 
where the last equality makes use of (P1) and (P2): in the sum over $m$, according to (P2), all $\hat{\mathcal{H}}_m$ with $m\geq 3$ commute with $\hat{\mathcal{H}}_1$ and can thus act directly on $|\underline{0} \rangle$, giving zero according to (P1).

We now introduce
\begin{equation}\label{eq:Bl} \hat{B}_l = \sum_{m=1}^{l} \hat{\mathcal{H}}_m, \end{equation} where, of course, $l$ is limited to $N_\mathcal{N}$, the total number of effective Hamiltonians. The operator $\hat{B}_l$ contains all the $\hat{\mathcal{H}}_m$ for $m \leq l$. Trivially,
\begin{equation}
\label{eq:bn+1}\hat{B}_{l+1} = \hat{B}_l + \hat{\mathcal{H}}_{l+1},
\end{equation}
and all $\hat{B}_l$ are hermitian since they are defined as sums of hermitian operators. We recall that the system's Hamiltonian is included in $\hat{\mathcal{H}}_1$, see Eq.~(\ref{eq:mathH1}), and thus included in all the operators $\hat{B}_l$, $l=1,\ldots,N_{\mathcal{N}}$. With this new notation we can write Eqs.~(\ref{eq:1}) and (\ref{eq:2}) as
\begin{equation}
\label{eq:12}|\underline{1} \rangle= \hat{B}_1 |\underline{0} \rangle;\quad |\underline{2} \rangle = \hat{B}_2 |\underline{1} \rangle.
\end{equation}

We now prove by induction that $|\underline{l+1} \rangle = \hat{B}_{l+1} |\underline{l} \rangle$.
Let us suppose by hypothesis that 
\begin{equation}
\label{eq:n2}|\underline{l} \rangle = \hat{B}_l |\underline{l-1} \rangle,
\end{equation} 
and evaluate $|\underline{l+1} \rangle$. Using Eq.~(\ref{eq:n}), one finds
\begin{eqnarray}
|\underline{l+1} \rangle &=& \sum_{m=1}^{N_\mathcal{N}} \hat{\mathcal{H}}_m |\underline{l} \rangle\\
&=& (\sum_{m=1}^{N_\mathcal{N}} \hat{\mathcal{H}}_m) \hat{B}_l |\underline{l-1} \rangle,
\end{eqnarray} 
where the sum runs over all the multiplets, {\em i.e.}, the full Hamiltonian is considered. However, in the summation, the $\hat{\mathcal{H}}_m$ with $m\geq l+2$ do not contribute to $|\underline{l+1}\rangle$ because $\hat{B}_l$ contains only the $\hat{\mathcal{H}}_k$ with $k\leq l$ and because of the properties (P1) and (P2). Consequently, only the terms $\hat{\mathcal{H}}_m$ with $m\leq l+1$ in the sum contribute and, therefore,
\begin{eqnarray}
|\underline{l+1} \rangle &=& ( \sum_{m=1}^{l+1} \hat{\mathcal{H}}_m ) \hat{B}_l |\underline{l-1} \rangle\\
 &=& ( \sum_{m=1}^{l+1} \hat{\mathcal{H}}_m )|\underline{l} \rangle\\
 &=& \hat{B}_{l+1} |\underline{l} \rangle,\end{eqnarray}
where we have used Eq.~(\ref{eq:Bl}). Thus, if the hypothesis Eq.~(\ref{eq:n2}) is true for $|\underline{l} \rangle$, it is true for $|\underline{l+1} \rangle$. Finally, since we have already shown that the hypothesis is valid for $l=1,2$ (see Eq.~(\ref{eq:12})), we have proven its validity for all $l$. Consequently, we immediately get:
\begin{equation}
\label{eq:bnn}|\underline{l} \rangle = \hat{B}_l \hat{B}_{l-1} \cdots \hat{B}_2 \hat{B}_1 |\underline{0} \rangle. \end{equation} 
With this result we can now evaluate $\mathcal{M}_{2n}$ and $\mathcal{M}_{2n+1}$ from Eqs.~(\ref{eq:M2n}) and (\ref{eq:M2n+1}):
\begin{eqnarray}
\mathcal{M}_{2l}   &=& \langle \underline{0} | \hat{B}_1 \cdots \hat{B}_{l-1} \hat{B}_l^2 \hat{B}_{l-1} \cdots \hat{B}_1 |\underline{0} \rangle\\
\mathcal{M}_{2l+1} &=& \langle \underline{0} | \hat{B}_1 \cdots \hat{B}_{l} \hat{B}_{l+1} \hat{B}_{l} \cdots \hat{B}_1 |\underline{0} \rangle.\end{eqnarray}

To obtain our final result, one last step is required for $\mathcal{M}_{2l+1}$. This quantity can be reduced to
\begin{eqnarray}
\nonumber \mathcal{M}_{2l+1} &=& \langle \underline{l} |\underline{l+1} \rangle = \langle \underline{l} | \hat{B}_{l+1} |\underline{l} \rangle\\
\label{eq:M2l+1_2}&=& \langle \underline{l} | (\hat{B}_{l} + \hat{\mathcal{H}}_{l+1}) |\underline{l} \rangle = \langle \underline{l} | \hat{B}_{l} |\underline{l} \rangle\end{eqnarray} because of
\begin{equation} \label{eq:av0}
\langle \underline{l} | \hat{\mathcal{H}}_{l+1} |\underline{l} \rangle = 0.
\end{equation}
To prove the last equality, one needs to explicitly consider the state $|\underline{l} \rangle$ in order to evaluate expectation values. Introducing explicitly the electronic basis, this state can be defined as
\begin{equation}
\label{eq:no} |\underline{l} \rangle = (|l^{(1)} \rangle, |l^{(2)} \rangle, \ldots, |l^{(N)} \rangle)^T,
\end{equation} where the superscript $(i)$ labels the component of the state $|\underline{l} \rangle$ in the corresponding electronic state $i$, and  
\begin{equation}
|l^{(i)}\rangle=|f^{(i)}_{S1\cdots l}\rangle|0_{l+1}\rangle\cdots |0_{N_\mathcal{N}}\rangle,
\end{equation} where $|f^{(i)}_{S1\cdots l}\rangle$ depends on the modes of the system and on the first $l$ multiplets of effective modes. Inspection of Eq.~(\ref{eq:bnn}) shows that the system's modes and the multiplets $k\leq l$ are inseparable in $|\underline{l} \rangle$ as indicated by $|f^{(i)}_{S1\cdots l}\rangle$ while the multiplets $k \geq l+1$ remain unaffected when constructing $|\underline{l} \rangle$. $\hat{\mathcal{H}}_{l+1}$ depends only on the multiplets $l+1$ and $l$, see Eqs.~(\ref{eq:Hm}) and (\ref{eq:mathH1}). We write $\hat{\mathcal{H}}_{l+1}$ as $\Omega_{l+1} (P_{l+1}^2 + X_{l+1}^2)/2 -\mathcal{E}_{l+1} + d_{l,l+1} (P_l P_{l+1} + X_l X_{l+1})$ where the indices refer to the $\mathcal{N}$ modes of the corresponding multiplets $l$ and $l+1$. Evaluating the expectation value $\langle \underline{l} | \hat{\mathcal{H}}_{l+1} |\underline{l} \rangle$ using $|\underline{l} \rangle$ in Eq.~(\ref{eq:no}), we readily see that all terms vanish: the harmonic oscillator contribution is cancelled by $\mathcal{E}_{l+1}$, and the bilinear part does not constribute because the expectation values of $X_{l+1}$ and $P_{l+1}$ in $|0_{l+1}\rangle$ are zero. This concludes the proof of Eq.~(\ref{eq:M2l+1_2}).

We now have
\begin{eqnarray}
\label{eq:finalM2n}\mathcal{M}_{2l} &=& \langle \underline{0} | \hat{B}_1 \cdots \hat{B}_{l-1} \hat{B}_l^2 \hat{B}_{l-1} \cdots \hat{B}_1 |\underline{0} \rangle,\\
\label{eq:finalM2n+1}\mathcal{M}_{2l+1} &=& \langle \underline{0} | \hat{B}_1\cdots \hat{B}_{l-1} \hat{B}_{l}^3 \hat{B}_{l-1} \cdots \hat{B}_1 |\underline{0} \rangle\end{eqnarray}
which proves that {\em only} the system's Hamiltonian augmented by the members $\hat{\mathcal{H}}_{m}$ with $m\leq l$ of the hierarchy of effective Hamiltonians are needed to exactly evaluate $\mathcal{M}_{2l}$ and $\mathcal{M}_{2l+1}$ and all the lower-order partial moments $\mathcal{M}_{k}$ with $k < 2l$. In turn, this implies that {\em by using the truncated hierarchy $\hat{H}_S + \sum_{m=1}^{n} \hat{H}_{m}$ we reproduce exactly all the moments $M_k$ with $k\leq 2n+1$}, see Eq.~(\ref{eq:Mn}). This proves the theorem and is the central result of this paper. As an immediate consequence one finds that each time one takes into account an additional member of the hierarchy, two additional moments of the macrosystem are recovered exactly. Further properties of the hierarchy are discuss below. 
 
\subsection{Discussion}

The hierarchy of effective Hamiltonians allows us to express the original Hamiltonian of the environment in terms of {\em sequential couplings of multiplets of effective modes} and is closely related to dynamical and spectral properties of the entire macrosystem. Using a given number of multiplets, the quantum dynamics on a given time-scale is numerically {\em exactly} reproduced, {\em whatever} the number of environmental modes is. This paves the way for studying truly large macrosystems involving a manifold of coupled electronic states. 

On the very-short time-scale, only the first member of the hierarchy plays a role. It the only member of the hierarchy which couples directly the electronic states as does the system's Hamiltonian $\hat{H}_S$. All the impact of the environment onto the system on this very-short time-scale is contained in this first member which determines the three first moments of the macrosystem. Then, at later times also the second member becomes relevant. Its effect is to spread the vibrational energy whiting each electronic state. There is no direct energy transfer between the system and the effective modes of the second member of the hierarchy. This transfer is mediated by the first member. At even later times the third multiplet comes also into play. This multiplet now spreads further the energy of the macrosystem, and so on. Whenever a new multiplet comes into play, the energy is further spread into more directions in the environment.

\subsubsection{Autocorrelation function and band shape of macrosystems}

Due to the large number of environmental modes and the presence of several coupled electronic states, the autocorrelation function, which measures the overlap between the initial wavefunction and the time-evolving one, decreases usually very rapidly. This is a very well known feature for 2-state conical intersections, where the autocorrelation function typically decays in the 10-100 femtosecond time-scale, and subsequently exhibits some oscillations \cite{worth:2004,domcke:2004}. These oscillations are usually of small amplitude because of the multimode nature of the dynamics which spread the wavefunction into many directions and lowers substantially the overlap with the initial one. The initial decay plays a crucial role. After this initial decay, the autocorrelation function typically possesses small values. This is of central interest here, since a few members of the hierarchy suffice to reproduce numerically exactly the short-time dynamics, and thus the ultrafast initial decay of the autocorrelation function. 

By Fourier transforming the autocorrelation function, one obtains the spectrum of the macrosystem. The above remarks on the autocorrelation function can be translated in the frequency domain as follows. The fast initial decay of the autocorrelation function determines the band shape of the spectrum, which is thus accurately reproduced by using a limited number of multiplets. The autocorrelation function at later times after the initial decay determines the fine structures of the spectra, carving the band-shape of the spectra. The typically small value of the autocorrelation function after the initial decay translates in small carving of the band-shape. Again, this is well known for 2-state conical intersections, where the spectra are often so dense that individual spectral lines are difficult to resolve even for relatively small molecular species \cite{worth:2004,domcke:2004}. By using the proposed approach, one can thus obtain accurate band shapes of truly large macrosystems by employing a few multiplets of modes only. These band shapes can be compared to experimental results. 

\subsubsection{How many moments of the Hamiltonian are independent?}

We have proven above that the truncated hierarchy $\hat{H}_S+\sum_{m=1}^{n} \hat{\mathcal{H}}_{m}$ reproduces exactly all the moments of the entire macrosystem up to $M_{2n+1}$. Now, for the sake of discussion, let us assume that we consider the full hierarchy, {\em i.e.}, $n=N_{\mathcal{N}}$. One then gets exactly all the moments up to $M_{2N_{\mathcal{N}}+1}$. On the other hand, since the hierarchy is complete we have taken account of the full Hamiltonian, and this implies that {\em all} the moments are exact, and not only the $2N_{\mathcal{N}}+2$ first ones (starting with $M_0$). We may conclude that there are only $2N_{\mathcal{N}}+2$ "independent" moments of the full macrosystem. If one determines correctly the first $2N_{\mathcal{N}}+2$ moments using our systematic approach, it follows immediately that all the higher moments are also correct. Let us consider an example. Assume there are $N_B=$24 environmental modes in a $N=$3-state problem. Each multiplet of modes contains $\mathcal{N}=$6 modes (sextet), and there are $N_{\mathcal{N}}=$4 sextets in total. This means that there are only 10 "independent" moments for this problem: $M_0$ to $M_9$. All the higher-order moments are automatically reproduced exactly if these first 10 moments are exact. We mention that the "independence" of the moments is to be understood with respect to the environment. The moments also depend on the system part, which is fully included in our theory, and for which all the moments can be "independent" depending on the complexity of $\hat{H}_S$. 

We stress that the property of the moments found above is a consequence of the form of the hierarchy of effective Hamiltonians, and the definition of the multiplets of effective modes. If one uses instead the original form of the Hamiltonian of the environment, {\em all} the original modes must be included in order to compute even the first non-trivial moment $M_2$. And, of course, once all the modes are included in the calculation, all the moments of the Hamiltonian are correctly reproduced. This is, however, of little help since we cannot include all these modes in any realistic calculation. Consequently, even the width of the spectrum (related to $M_2$) cannot be reproduced exactly when using the original Hamiltonian of the macrosystem.

\subsubsection{Truncation of the hierarchy vs. truncation of the moment expansion of the autocorrelation function}

The following remark is of fundamental importance. Computing the quantum dynamics provided by the system's Hamiltonian augmented by the hierarchy truncated at the member $n<N_{\mathcal{N}}$, one reproduces exactly the $2n+2$ first moments ($M_0$ to $M_{2n+1}$) of the full Hamiltonian of the macrosystem. We stress that {\em the quantum dynamics provided by a truncated hierarchy does not lead whatsoever to a truncation of the Taylor expansion of the autocorrelation function} -- which would have been dramatic for the dynamics on a longer time-scale. This can be easily seen. First, let us recall the definition of the moments, see Eq.~(\ref{eq:MM}), using the truncated hierarchy at the order $n$ instead of the full Hamiltonian. The first moment which is not exactly reproduced by this approximate Hamiltonian is $M_{2n+2}$ and reads 
\begin{equation} \label{eq:M2n+2}
M_{2n+2} = \langle \underline{0} | (\hat{H}_S + \sum_{m=1}^{n}\hat{H}_m)^{2n+2} | \underline{0} \rangle.
\end{equation} Obviously, this moment does {\em not} vanish. Only the contributions arising from the first neglected term $H_{n+1}$ of the hierarchy are missing in Eq.~(\ref{eq:M2n+2}) in order to make this moment exact too. Similar arguments hold for all other higher-order moments. Therefore, we not only obtain exactly the $2n+2$ first moments, but also important contributions to {\em all} the higher-order moments. The results on the dynamics at times longer than those reproduced accurately by the $2n+2$ first moments are approximate, but do contain the impact of important contributions of all the higher moments. 

\subsubsection{The wavefunction of the macrosystem}

As time proceeds, the propagation of the exact wavefunction depends on an increasingly growing number of members of the hierarchy of effective Hamiltonians. Let us call $t_n$ the time up to which the dynamics is accurately reproduced by employing only the $n$ first members of the hierarchy. In order to propagate the wavefunction, we consider in a dynamical calculation only the part of the wavefunction which depends on the first $n$ multiplets we want to account for. The remaining part needs not to be explicitly propagated and evolves trivially as free oscillators. Consequently, even when using a truncated hierarchy, we {\em know the full-dimensional wavefunction of the entire macrosystem} and all quantum observables can be, in principle, evaluated.

We do not have a mathematical criterium to define {\em a priori} the value of $t_n$ up to which the dynamics is accurately reproduced by employing the truncated hierarchy of effective Hamiltonians. Note that this value will depend also on the system's parameters. However, it is possible to estimate this value numerically by using the following procedure. We know that when we add a member of the hierarchy, two more moments of the full Hamiltonian are exactly reproduced. Thus, if we compare the numerical results (autocorrelation functions) computed using $n$ members to those obtained by using $n+1$ members, the results will start to deviate at $t_n$. In fact, it is sufficient to add a single mode of the multiplet number $n+1$, and compare the two autocorrelation functions obtained with and without this additional mode in order to get an estimate of $t_n$. The wavefunction computed for times later than $t_n$ will be of diminishing quality as time increases, but will nevertheless account for important contributions of the exact wavefunction. This is easily seen from the discussion in point 3 of this section.

\subsubsection{Effective modes vs. normal modes}

Our last point in this discussion concerns the link between the complete set of $N_B$ effective modes and the original normal modes. To construct the effective modes, we have applied successive orthonormal transformations (rotations) of the original normal modes, see Sec. III. These rotations can, of course, be concatenated into a single one, which we call $\bm{T}$, and which is nothing else than the product of all the successive rotation matrices. Applied to the original modes $\bm{x}$, this single rotation leads directly to the complete set of effective modes $\bm{X}$ which enter the hierarchy of effective Hamiltonians: $\bm{X}=\bm{T}\bm{x}$. It is obvious that, by using the transposed transformation $\bm{T}^T$, we recover the original modes out of the effective modes: $\bm{x}=\bm{T}^T\bm{X}$.

Since the full-dimensional wavefunction is known numerically exactly up to $t_n$, see the last point discussed above, highly accurate expectation values can be computed in the full space of effective modes. On the other hand, knowing the transformation matrix from this space to the original space of normal modes, one can compute expectation values of quantities defined by the latter modes. Examples are the time-dependent energy gained by some normal modes of interset and the average position of a particular normal mode. Such quantities are of help to interpret experimental results, where discussions in terms of normal modes are widely used. 

\section{Conclusion}

The Hamiltonian of the macrosystem is decomposed into a system part and an environment part. The latter is described by the LVC model and the former is comprised of a few modes which are not restricted to any model and usually include the modes which cannot be satisfactorily described by the LVC model. It is the huge number of environmental modes which exclude a direct, numerically exact calculation of the quantum dynamics of the macrosystem.

Starting from the original Hamiltonian of the environment we derived a hierarchy of effective Hamiltonians. Each member of the hierarchy of effective Hamiltonians has been shown to depend on a single multiplet of effective modes. The number of effective modes within each multiplet depends on the number of coupled electronic states. The first member of the hierarchy plays a key role since it contains all the direct coupling among these electronic states due to the environment. Only this first member participates directly, together with the system part, in the coupling of the electronic states. The higher members of the hierarchy are all diagonal in the electronic space, {\em i.e.}, they do not couple the manifold of electronic states.

Furthermore, except of the first member, each member of the hierarchy is coupled only to the "former" member by bilinear kinetic and potential terms. As a consequence, the original Hamiltonian of the environment where all the modes formally play a similar role, is replaced by a hierarchy of Hamiltonians which exhibits a {\em sequential coupling of multiplets of modes}. This sequential coupling immediately translates to a hierarchical description of the quantum dynamics of the full macrosystem. We have proven, by analyzing the moments of the autocorrelation function of the entire macrosystem, that each member of the hierarchy comes into play at a different time. The truncated hierarchy at the order $n$ suffices to reproduce exactly the first $2n+2$ moments of the entire macrosystem. The higher moments are also reproduced but only approximately. In this way one obtains numerically exactly the dynamics of the macrosystem up to a finite time $t_n$ which grows as $n$ increases. The total wavefunction of the macrosystem is known in the full space of vibrational modes. Thus, apart from the autocorrelation functions, spectra, and the time-evolving populations of the electronic states, we can also evaluate other quantities related to the wavepacket, like, for instance, the individual excitation along all normal modes.

This work extends the 2-state effective Hamiltonian formulation for short-time dynamics \cite{prl:2005} to the more general case of $N$ coupled electronic states and lifts the restriction to short-time dynamics. We expect this work to be valuable in studying the highly complex behavior of electronically excited macrosystems involving multi-state intersections and a large number of nuclear modes. The use of a truncated hierarchy, {\em i.e.}, of a reduced number of effective modes for the environment, can allow for a numerically exact treatment of the quantum dynamics in such macrosystems. Finally, we mention that the hierarchy of effective Hamiltonians is certainly of interest also for the use of approximate dynamical methods. 

\begin{acknowledgments}
Fruitfull discussions with I. Burghardt and H. K{\"o}ppel are greatfully acknowledged. This work has been supported financially by the Deutsche Forschungsgemeinshaft (DFG).
\end{acknowledgments}


\end{document}